# Effect of kick velocity on the gamma-ray burster distribution

**Nobuo TERASAWA**[1] **and Makoto HATTORI**[2]

[1]Institute of Physical & Chemical Research (RIKEN)
Hirosawa 2-1, Wako, Saitama 351-01, Japan

[2]Max-Planck-Institut für extraterrestrische Physik
Postfach 1603, D-85740 Garching bei München, Germany

## ABSTRACT

The effect of kick velocity to newly born pulsars on the distribution of $\gamma$-ray bursters is examined in the context of a disk origin model and a halo model of $\gamma$-ray bursters. The conversion formula from a two-dimensional velocity distribution function to a three-dimensional distribution function is derived and applied to reproduce the distribution function of the kick velocity of radio pulsars. Monte Carlo simulations of the kicked neutron stars show that the disk neutron star model of $\gamma$-ray bursters still needs unnatural assumptions if the velocity distribution of the $\gamma$-ray bursters is the same as that of neutron stars—only neutron stars with very high kick velocities can become $\gamma$-ray bursters and there are silent majorities. On the other hand, the core radius of $\gamma$-ray bursters is not found to be extended by the kick velocity if the core-halo structure similar to the Galactic dark matter distribution is based on the initial distribution of the halo neutron stars. Thus the introduction of a kick velocity to neutron stars do not improve the statistics of neither the disk model nor the halo model. Two possibilities to save Galactic models are suggested: (1) The $\gamma-$ray bursters are old neutron stars which were accelerated to velocities faster than 750 km s$^{-1}$ by jet propulsion and passed the death line for pulsars due to spin down from the rotational energy loss by jet ejection. (2) The initial distribution of neutron stars is fairly uniform and the extent of the halo is large enough for the halo model to be consistent with the observations. It implies that the initial star formation burst of the Galaxy occurred fairly uniformly in the extended halo region.

[1]e-mail: terasawa@riken.go.jp
[3]e-mail: mhattori@rosat.mpe-garching.mpg.de

## 1 Introduction

Though many models have been proposed for $\gamma$-ray bursters, they are essentially classified into two categories—cosmological models and Galactic models. The extreme isotropy of bursters observed by BATSE favors cosmological models. However, GINGA observations of cyclotron absorption lines, which are unique reliable clues to solve the nature of the $\gamma$-ray bursters (Murakami et al. 1988, Fenimore et al. 1988, and Yoshida et al. 1991), strongly support Galactic models since these same features have been detected in Galactic X-ray pulsars (e.g. Trümper et al. 1978, Makishima & Mihara 1992). The difficulty in the Galactic model is to reproduce both the observed isotropy and the non-uniform distribution of the bursters implied by $V/V_{max}$. Recently it has been reported that radio pulsars are born with very high space velocities (Lyne & Lorimer 1994). If old neutron stars are the source of $\gamma$-ray bursters and neutron stars generally have high kick velocities as implicated by the radio pulsar observations, the Galactic origin model may be saved since the distribution of neutron stars can be extended by their kick velocities. In order to examine the effect of the kick velocity on the distribution of bursters, we perform Monte Carlo simulations of the distribution of kicked neutron stars on the basis of the Galactic model of $\gamma$-ray bursters.

First of all, the conversion formula from the observable two-dimensional velocity distribution function in a projected space to the three-dimensional velocity distribution function in a real space is derived and then applied to the kick velocity distribution of radio pulsars. The orbits of kicked neutron stars in the Galactic potential are traced with a high precision and the various statistical values are evaluated for comparison with observations.

## 2 Distribution function of kick velocities to neutron stars and neutron star orbits in the Galactic potential

As far as the radio pulsars concerned, only their proper motions are observable and hence the transverse velocities of pulsars can be estimated. We require the three-dimensional velocity distribution function of radio pulsars in real space in order to incorporate the effects of kick velocities to neutron stars. First, we show how the velocity distribution function in real space is restored from the observationally available transverse velocity distribution function.

It is straightforward to show that, under the assumption of isotropy, the two-dimensional velocity distribution function $f_{2D}$ is related to the three-dimensional velocity distribution function $f_{3D}$ by the following Abel integral

$$f_{2D}(t) = t \int_t^\infty \frac{f_{3D}(v)}{v\sqrt{v^2 - t^2}} dv. \qquad (1)$$

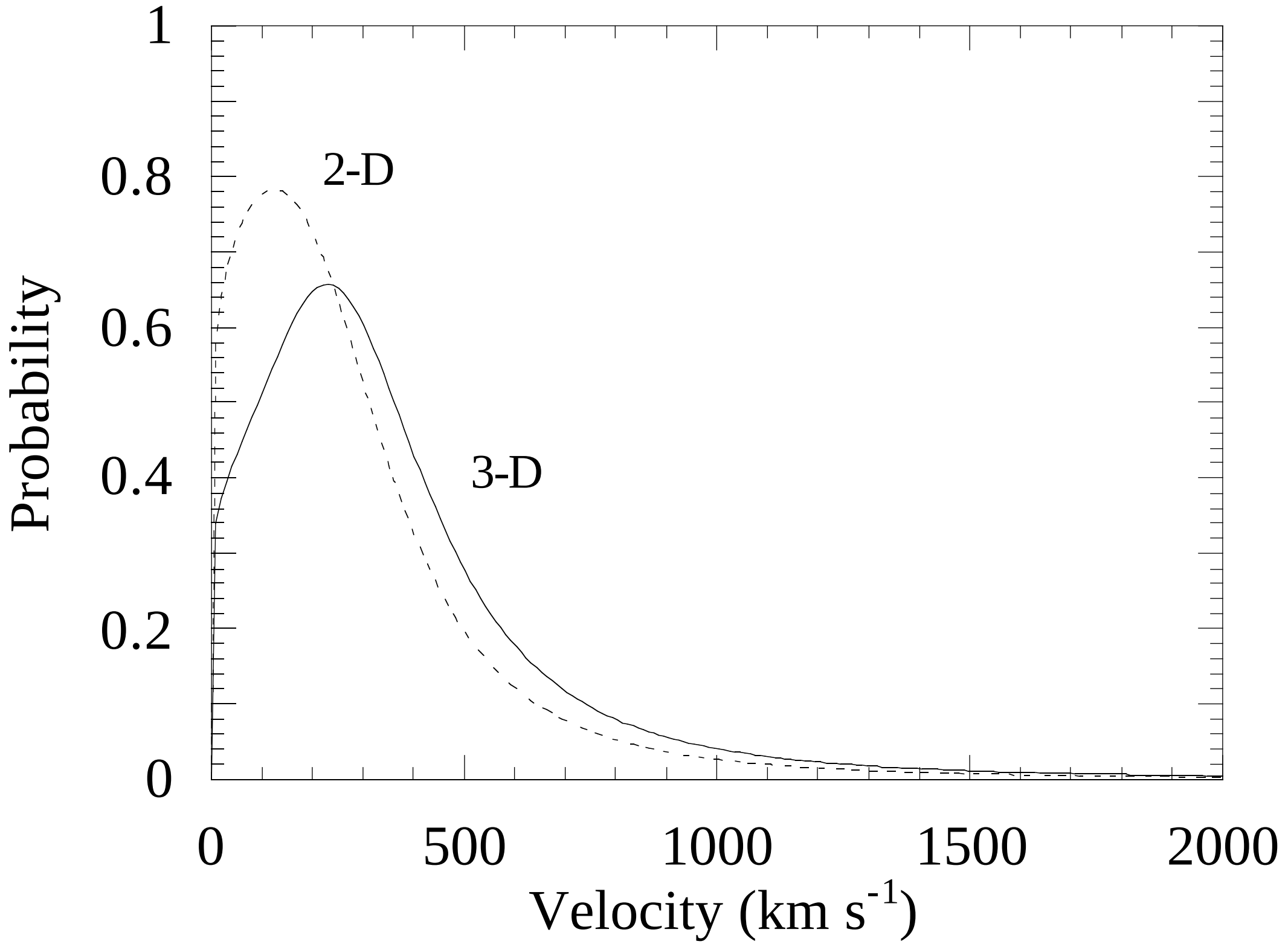

**Fig.1**. Three-dimensional velocity distribution function reconstructed from the analytical approximation for the two-dimensional velocity distribution of the radio pulsars.

This integral is inverted to give desired three-dimensional velocity distribution function by the Laplace transformation,

$$f_{3D}(v) = -\frac{2v^2}{\pi}\int_v^\infty \left(\frac{f_{2D}(t)}{t}\right)' \frac{dt}{\sqrt{t^2 - v^2}}, \tag{2}$$

where the prime denotes derivatives with respect to the transverse velocity $t$.

This formula is applied to reproduce the three-dimensional velocity distribution function of kick velocities of radio pulsars. Lyne and Lorimer (1994) found that the distribution of the observed two-dimensional transverse velocities of radio pulsars may be approximated by,

$$f_{2D}(t)dt \propto \frac{t^{0.13}}{1+t^{3.3}}, \quad t = \frac{v}{v_0}, \quad v_0 = 330 \text{ km s}^{-1}. \tag{3}$$

The three-dimensional velocity distribution function given by the transformation of eq. (??) is plotted in Fig. 1. The derived 3-D velocity distribution function has a mean of 410km $s^{-1}$ and r.m.s. value of 689km $s^{-1}$.

We examine the effect of kick velocities to newly born neutron stars for three models of gamma-ray bursters—the Galactic disk origin model, the disk origin model with the assumption that only the high velocity neutron stars can be the $\gamma$-ray bursters, and the halo origin model.

Neutron stars originating in the Galactic disk are assumed to be born during the stages of initial star formation burst and/or born with a constant birth rate from the formation of the Galaxy to the present time. The distribution of the disk neutron stars, the Galactic potential, and the calculation procedure of the orbits of the neutron stars are essentially same as those employed by Paczyński (1990). An important difference here is that the kick velocities are supplied from our Monte Carlo method according to the velocity distribution function of radio pulsars.

The random velocities generated according to the 3-D velocity distribution function are added to the velocities of newly born neutron stars. Following the approach of previous authors (Burton & Gordon, 1978, Binney & Tremaine, 1987), we assume that the progenitors of neutron stars born in the Galactic disk have rotational velocities of the form adopted by Paczyński (1990). On the other hand, the progenitors in the Galactic halo are assumed to be at rest. This assumption is justified since the lifetime of massive stars as the progenitor of neutron stars is much shorter than the dynamical time of the halo.

The probability distribution for the location of the disk neutron stars is assumed to be of the form,

$$p_z dz = e^{-z/z_{exp}} \frac{dz}{Z_{exp}}, \quad (4)$$

$$p_R dR = a_R e^{-R/R_{exp}} \frac{R}{R_{exp}^2} dR, \quad (5)$$

$$a_R := [1 - e^{-R_{max}/R_{exp}}(1 + R_{max}/R_{exp})]^{-1} = 1.0683,$$

where z is the distance above the Galactic plane and R is the distance from the Galactic center. For the values of $z_{exp}$, and $R_{max}$, we adopt $z_{exp} = 75$ pc, $R_{max} = 4.5$ kpc (van der Kruit 1987).

The distribution of halo neutron stars is assumed to have a core-halo structure and hence the probability distribution is given as,

$$p_n(r)dr = \begin{cases} \frac{p_0}{1+(r/r_n)^2}\left(\frac{r}{r_{max}}\right)^2 dr & r \le r_{max} \\ 0 & r > r_{max} \end{cases} \quad (6)$$

where $r$ is the distance from the Galactic center. The values of the core radius, $r_n$, and the cut-off radius, $r_{max}$, of the neutron star halo are left as parameters in our simulations.

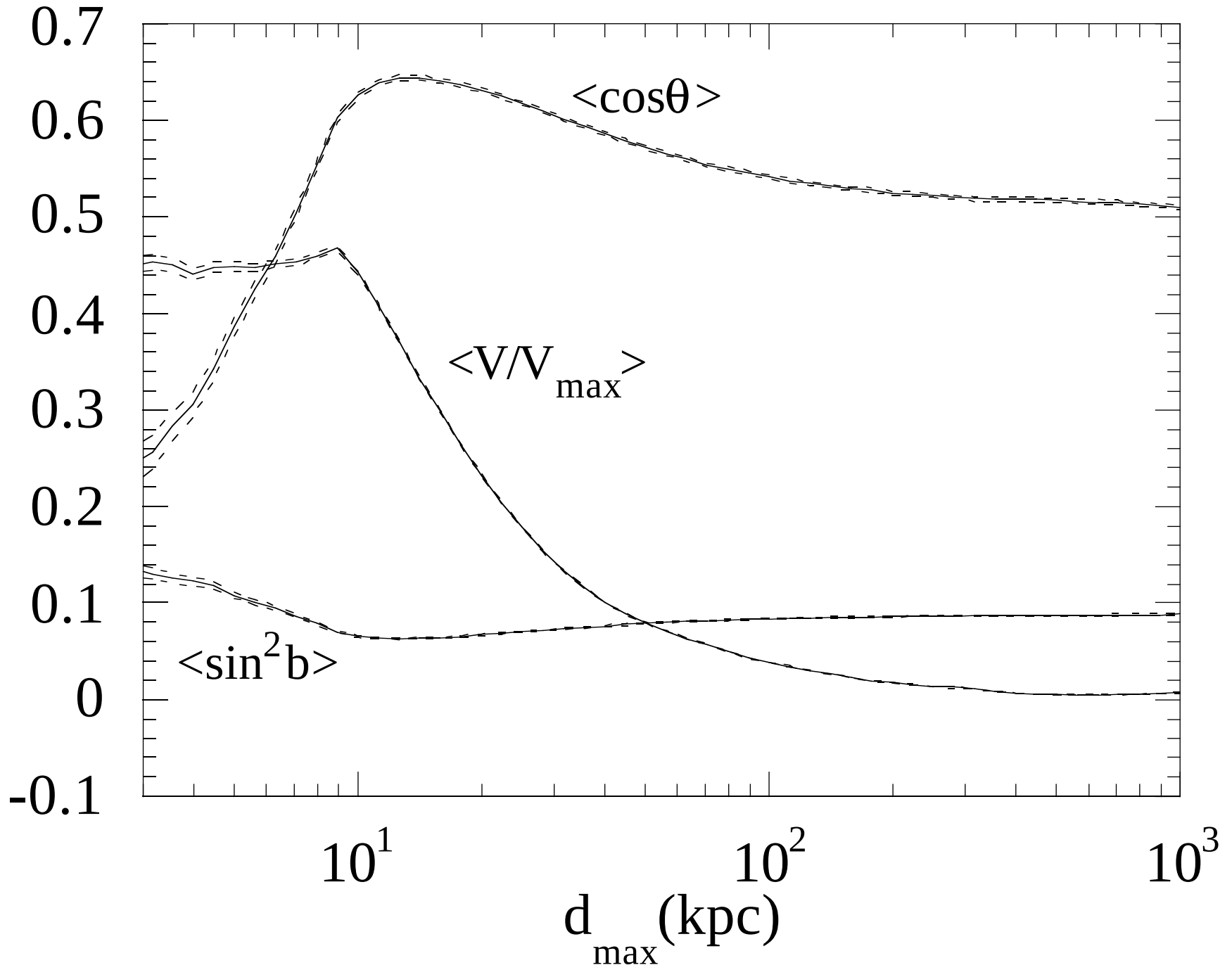

Fig. 2a

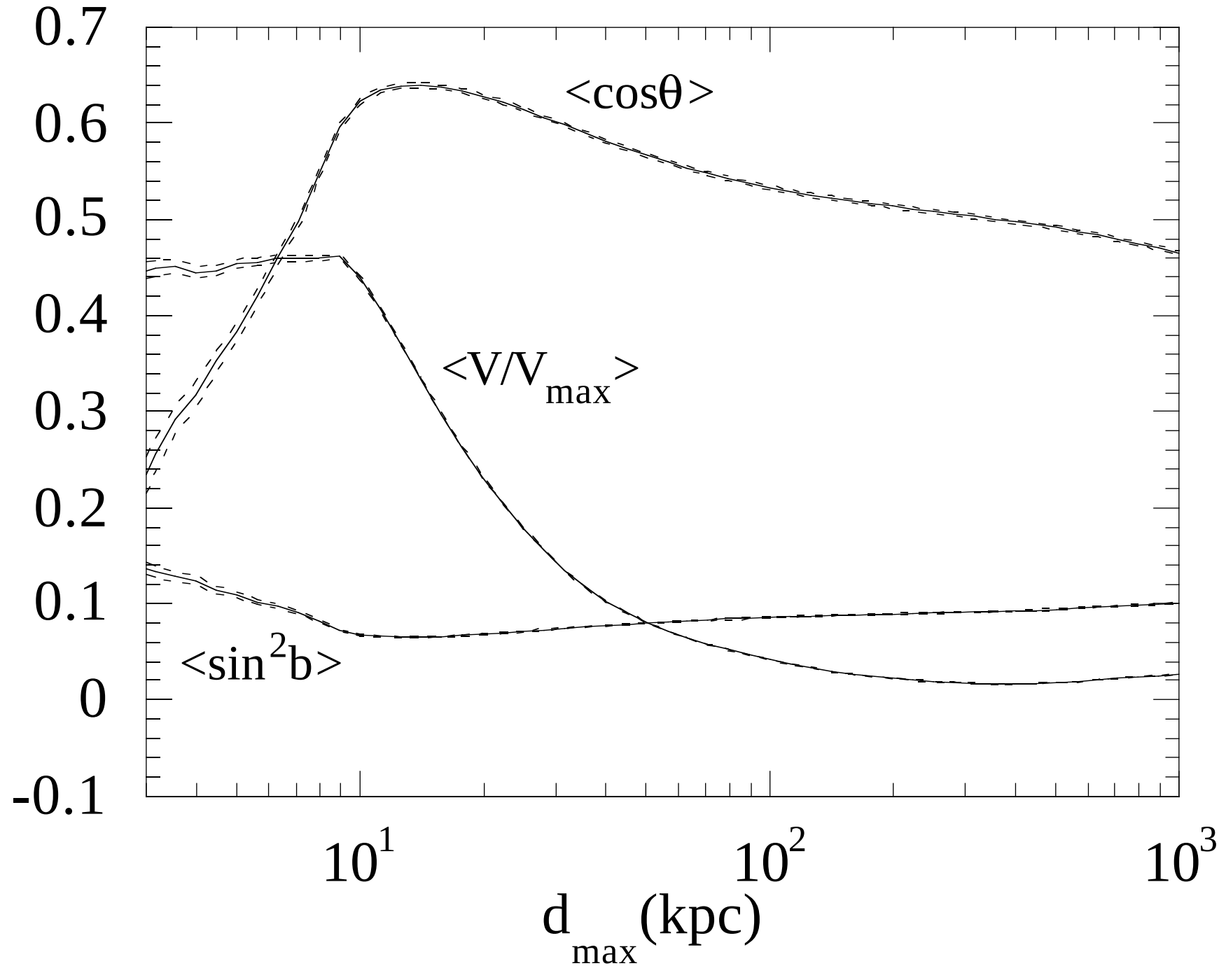

Fig. 2b

**Fig. 2a–b**. Dipole moment $< \cos\theta >$, quadrapole moment $< \sin^2 b >$, and $< V/V_{max} >$ of the $\gamma$-ray bursters plotted against the sampling distance $d_{max}$ in the disk origin model. The bursters are assumed to be born in the initial star formation burst (a) and born with a constant rate from the formation of the Galaxy to the present (b).

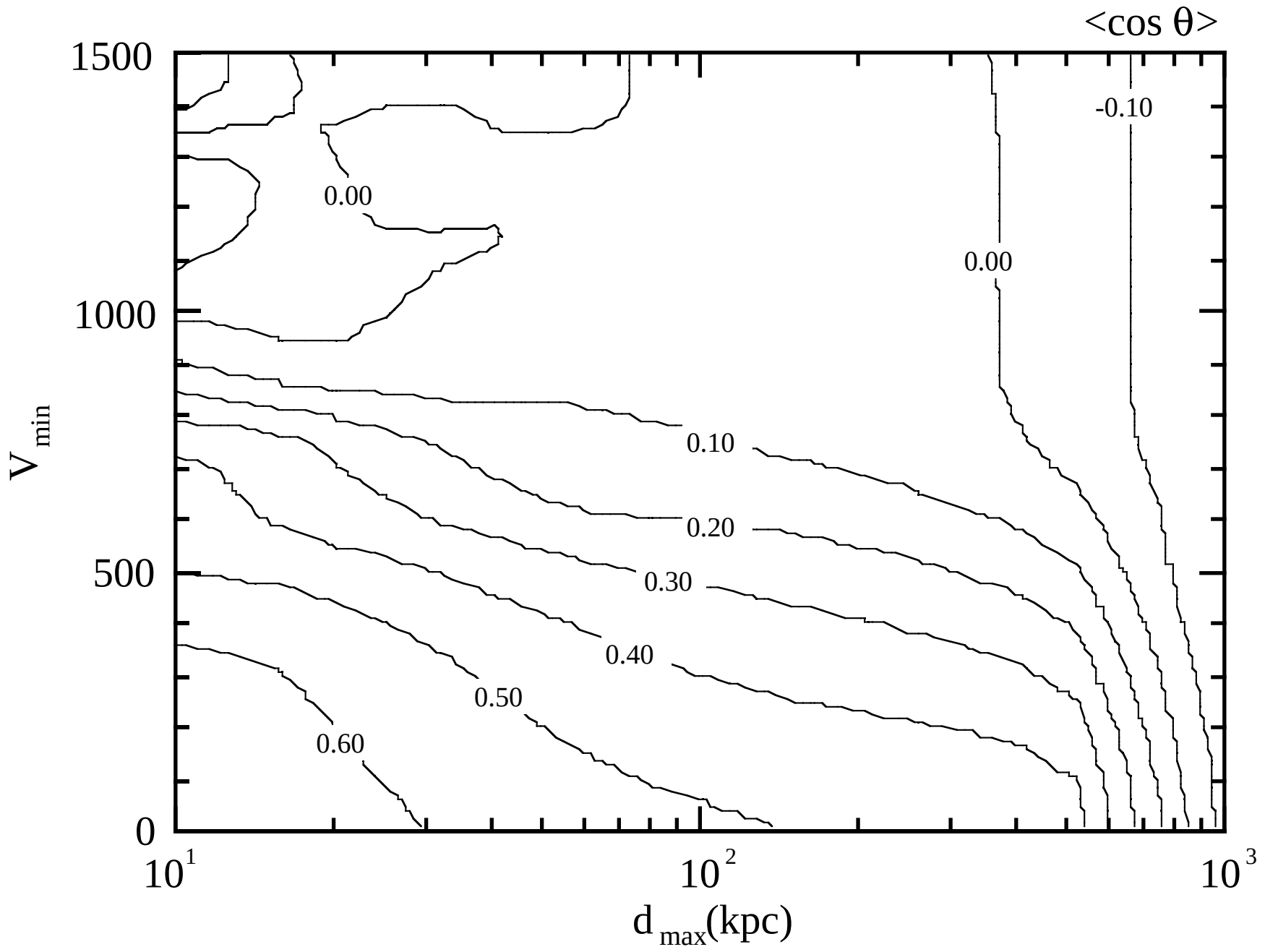

Fig. 3a

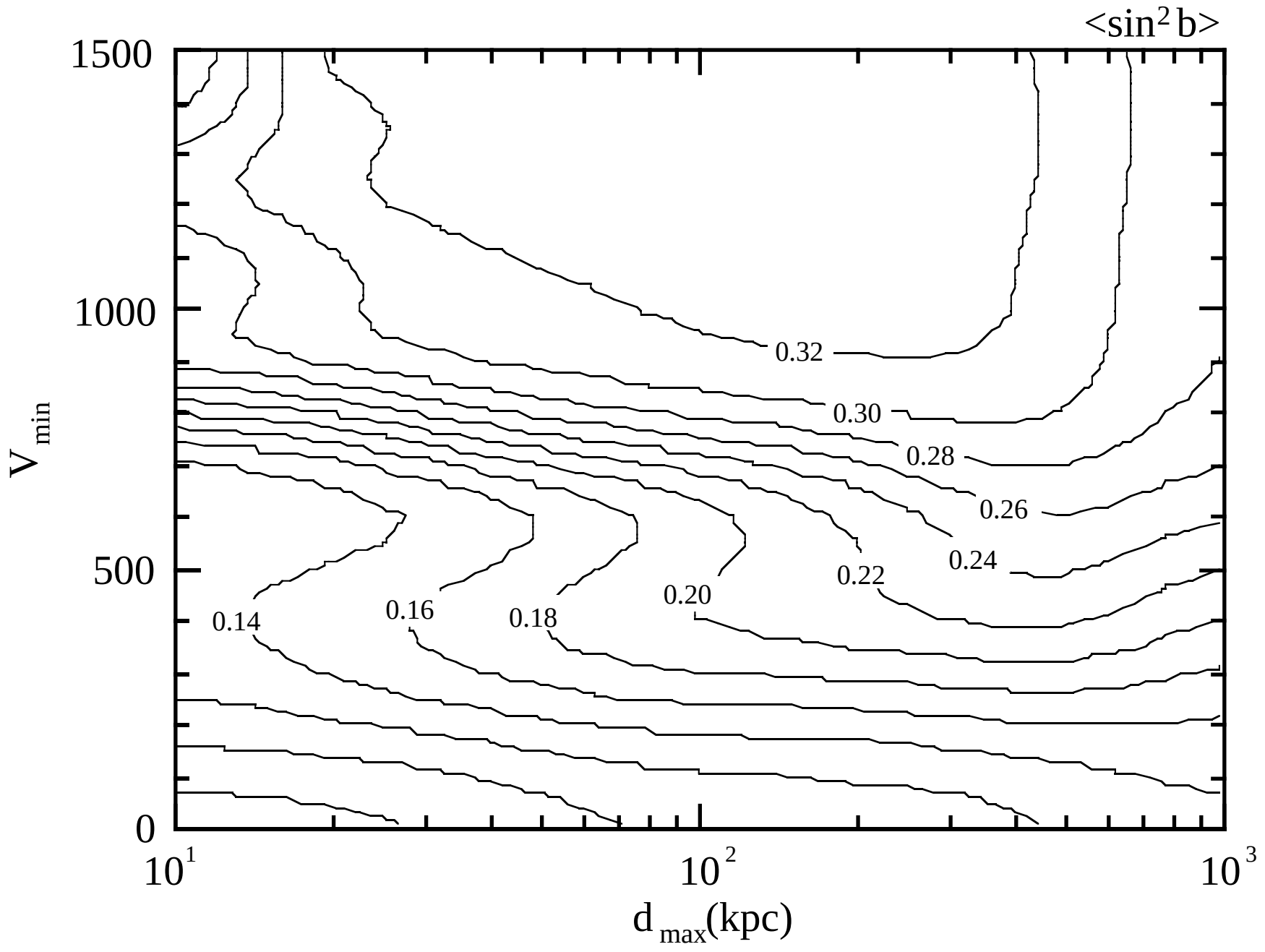

Fig. 3b

**Fig. 3a–d** Contour maps of the dipole moment (**a**), the quadrapole moment (**b**), the $< V/V_{max} >$(**c**), and 90% confidence level of the model (**d**), of the $\gamma$-ray bursters on the plane of the minimum velocity cutt-off $v_{min}$ and the sampling distance $d_{max}$ in the disk origin model. A constant birth rate and the bursting rate function proposed by Li and Dermer (1992) with the time scale of radio pulsars $\tau = 30$ Myr are assumed. In Fig. 3d, the solid line shows the 90% confidence level from $< \cos\theta >$ and $< \sin^2 b >$, and the broken line shows the 90% confidence level from $< \cos\theta >$, $< \sin^2 b >$, and $< V/V_{mx} >$.

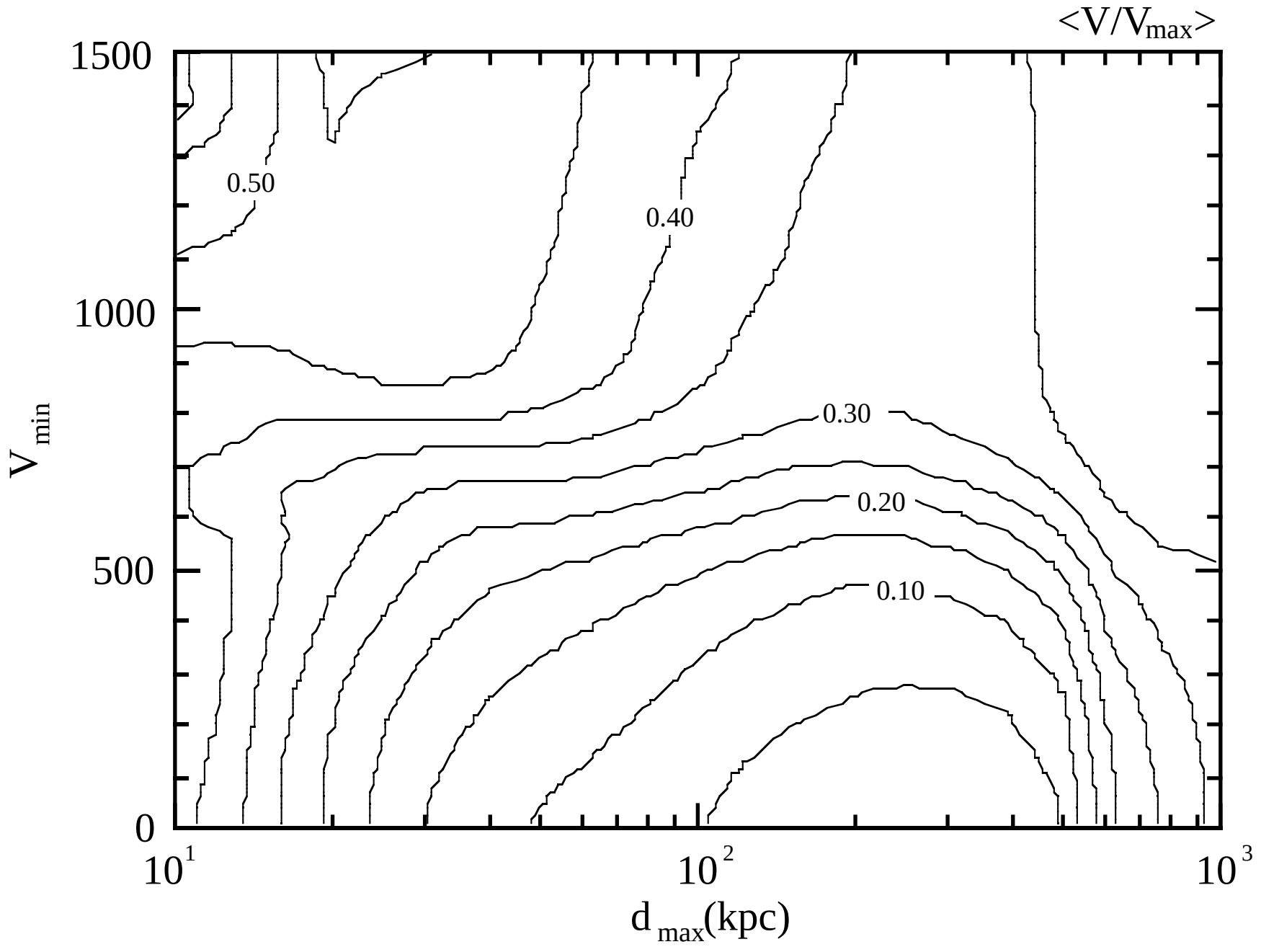

Fig. 3c

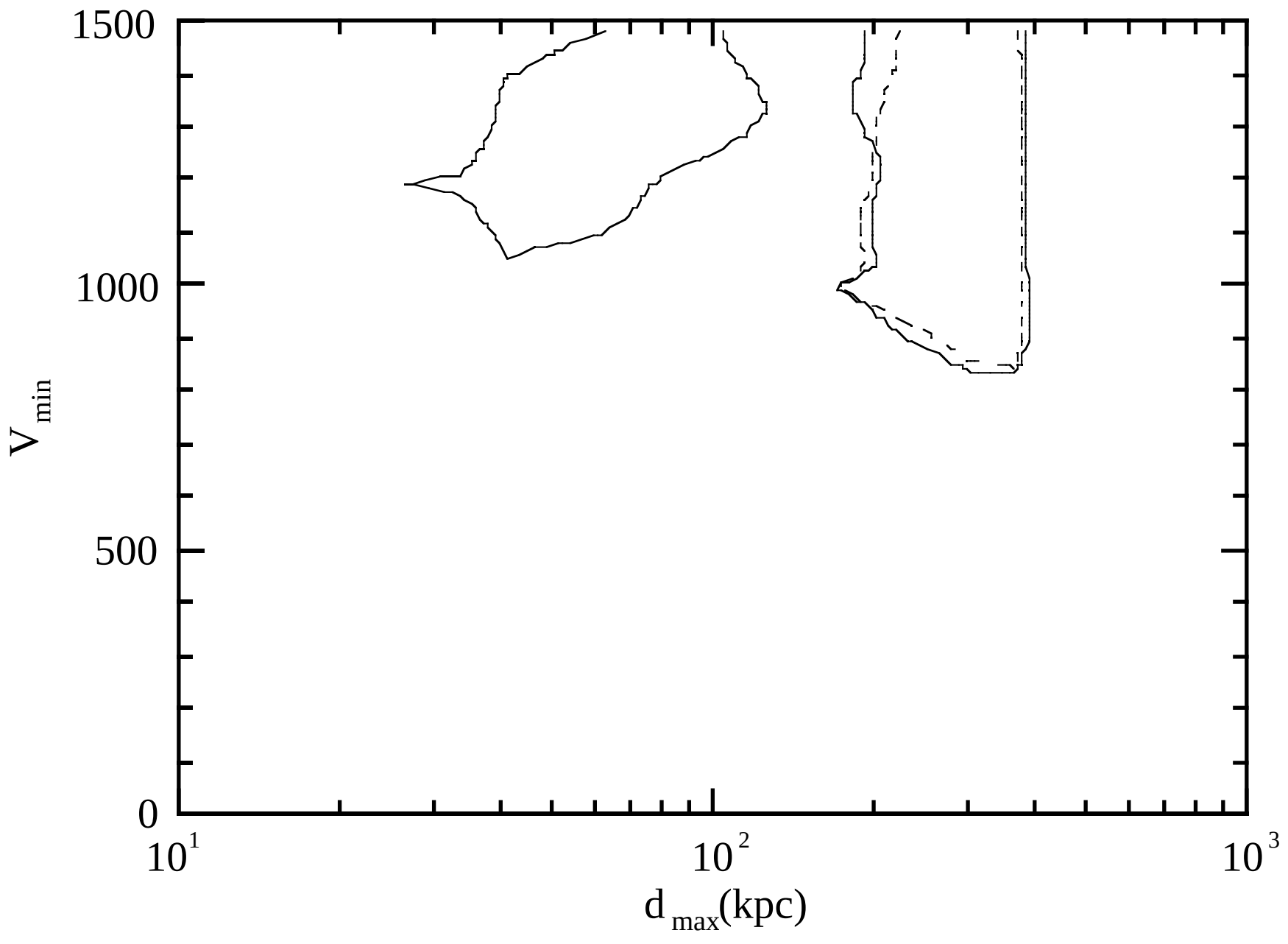

Fig. 3d

The Galactic graviational potential is assumed to be composed of three components,

$$\Phi_i(R,z) = \frac{GM_i}{\{R^2+[a_i+(z^2+b_i^2)^{1/2}]^2\}^{1/2}}, \tag{7}$$

$$\Phi_h = -\frac{GM_c}{r_c}\left[\frac{1}{2}\ln\left(1+\frac{r^2}{r_c^2}\right)+\frac{r_c}{r}\tan^{-1}\left(\frac{r}{r_c}\right)\right], \tag{8}$$

$$M_c := 4\pi\rho_c r_c^3, \quad r^2 = R^2+z^2,$$

where $i=1$ corresponds to the Galactic spheroid, $i=2$ corresponds to the Galactic disk (Miyamoto & Nagai 1975), and the halo component corresponds to the density distribution given as,

$$\rho_h = \begin{cases} \dfrac{\rho_c}{1+(r/r_c)^2} & r \le 70 \text{ kpc} \\ 0 & r > 70 \text{ kpc} \end{cases} \tag{9}$$

The parameters in the gravitational potential are chosen to be same as those adopted by Paczyński (1990),

$$a_1 = 0, \quad b_1 = 0.277 \text{ kpc}, \quad M_1 = 1.12\times10^{10}\ M_\odot, \tag{10}$$

$$a_2 = 3.7 \text{ kpc}, \quad b_2 = 0.20 \text{ kpc}, \quad M_2 = 8.07\times10^{10}\ M_\odot, \tag{11}$$

$$r_c = 6.0 \text{ kpc}, \quad M_c = 5.0\times10^{10}\ M_\odot. \tag{12}$$

The orbits of neutron stars are calculated numerically by integrating a set of equations of motion with a fourth-order Runge-Kutta method,

$$\frac{dR}{dt} = v_R, \quad \frac{dv_R}{dt} = \left(\frac{\partial\Phi}{\partial R}\right)_z + \frac{j_z^2}{R^3}, \tag{13}$$

$$\frac{dz}{dt} = v_z, \quad \frac{dv_z}{dt} = \left(\frac{\partial\Phi}{\partial z}\right)_R, \tag{14}$$

$$\Phi = \Phi_1+\Phi_2+\Phi_h,$$

where $j_z$ is angular momentum. These equations are integrated keeping the accuracy of the energy conservation better than $10^{-10}$.

A Monte Carlo method is employed to sample the initial positions of neutron stars and their random kick velocities. A total of $10^6$ orbits are traced for each model up to the assumed Galactic age of $10^{10}$yr. Since most of the high velocity neutron stars escape and few of them are left, $10^6$ neutron stars are added to the simulation for the samples with $v_{min} \geq 700$ km s$^{-1}$ and $v_{min} \geq 1000$ km s$^{-1}$.

The contributions from the Large Magellanic Cloud (LMC), the Small Magellanic Cloud (SMC), and M31 are incorporated simply by weighting the number of bursters with the luminous masses of these galaxies.

## 3 Result of Monte Carlo Simulation

First, we show the statistical results for the disk origin models (Fig. 2), which should be compared with the BATSE 3B data for 1122 bursts; $< \cos\theta >= -0.002 \pm 0.018$, $< \sin^2 b - 1/3 >= -0.003 \pm 0.009$, and $V/V_{max}$ for 657 bursts, $< V/V_{max} >= 0.33 \pm 0.01$ (Meegan et al. 1995). Both cases where 1) neutron stars are born in the initial star formation burst and 2) constant supernova explosions produce neutron stars at a constant rate, clearly contradict the statistical analysis of $\gamma$-ray bursters observed with BATSE if all neutron stars are potential $\gamma$-ray bursters. The dipole moment, $< \cos\theta >$, and the quadrupole moment $< \sin^2 -1/3 >$ are consistent with observations only in the very local region around the sun, say $r \lesssim 1$ kpc. However, the neutron star distribution is almost uniform and $< V/V_{max} >$ is too large to be consistent with the implication of non-uniform distribution of $\gamma$-ray bursters. Therefore another component is required for models where local neutron stars in the solar neighborhood are the source of $\gamma$-ray bursters (e.g. Smith & Lamb 1993).

If only high velocity neutron stars can be $\gamma$-ray bursters, as in the model proposed by Li and Dermer (1992), the statistics are improved and made consistent with the BATSE observation (Fig. 3). We find that we still need a minimum velocity cut-off, $V_{min} \gtrsim 750$ km s$^{-1}$ (Fig. 3d) in spite of the use of the new velocity distribution of kick-velocities if the velocity distribution of the $\gamma$-ray bursters is same as that of radio pulsars. The difficulty of this type of model is clearly that there is no rational physical mechanism to prevent low velocity neutron stars from becoming $\gamma$-ray bursters. Although a sharp cut-off in the kick velocity is needed for the burster mechanism, it seems difficult to realize such a cut-off from, for instance, the observed correlation between the velocity of pulsars and the magnetic field strength (see also the claim that this correlation is merely apparent due to the bias, Itoh & Hiraki 1994). Based on their recent observations, Markwardt and Ögelman (1995) suggest that the high velocity motion of radio pulsars may be the result of propulsion by the gas jet of pulsars based on their observation. They also suggest that the jet may result in the loss of the rotational energy of pulsars. If neutron stars, which are accelerated further from the pulsar stage and pass the so-called death line for pulsars due to the spin down caused by the rotational energy loss, turn out to be $\gamma$-ray bursters, the velocities of bursters can be much higher than those of the radio pulsars. If the majority of burster velocities are higher than 750 km s$^{-1}$, they can be consistent with

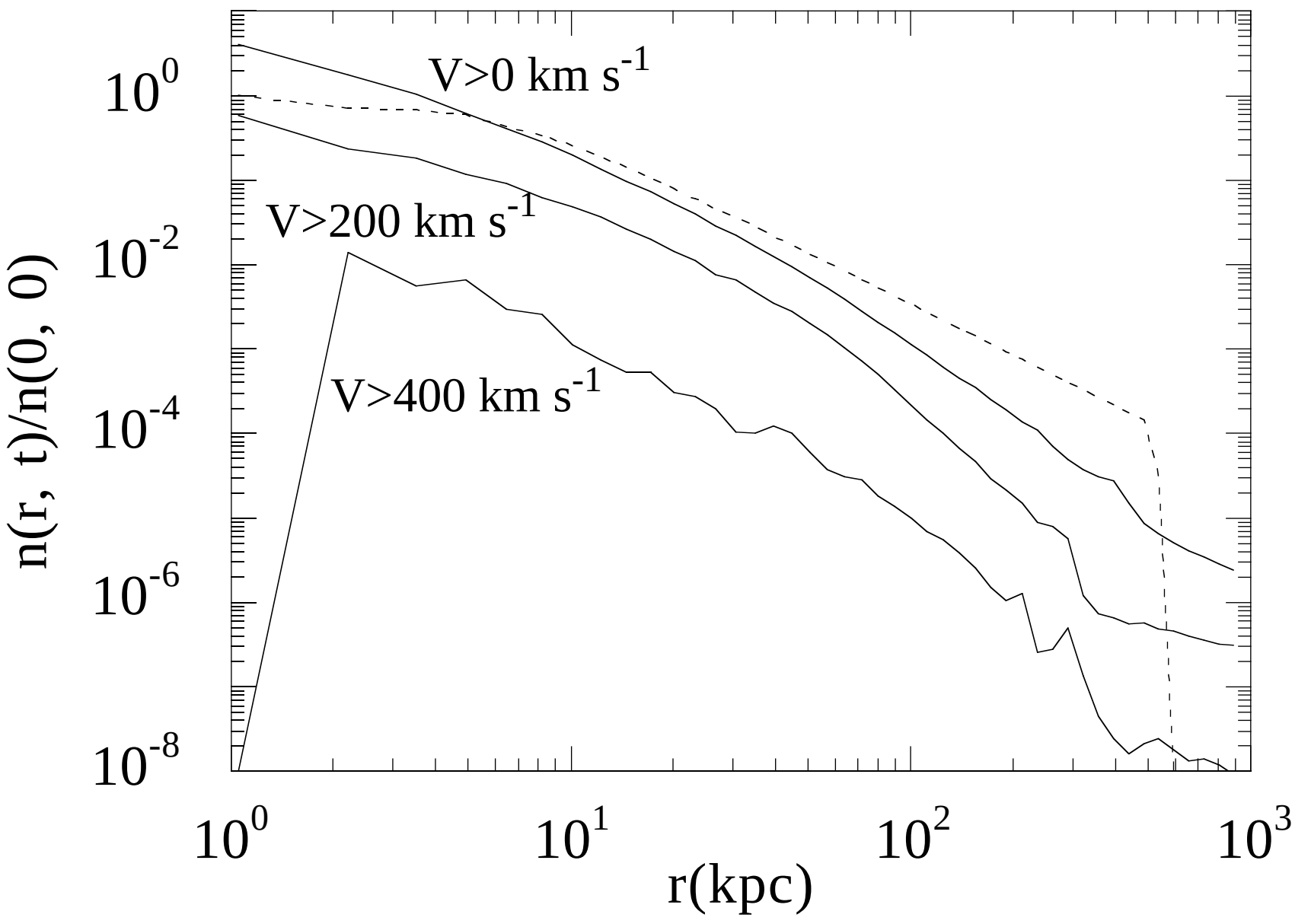

Fig. 4a

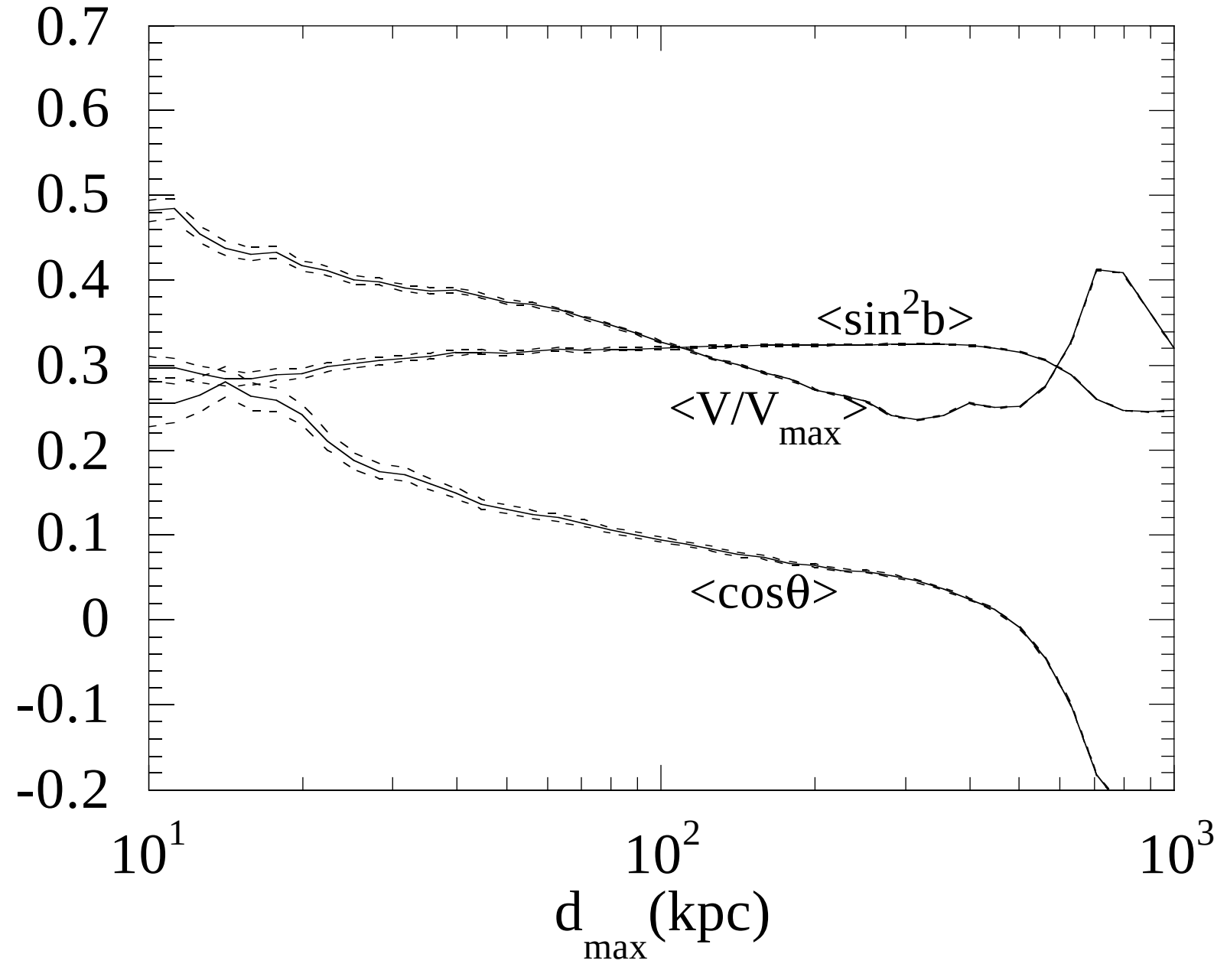

Fig. 4b

**Fig. 4 a–d** Results of the halo model: **a** The initial distribution (dotted line) and the final distribution (solid lines) after $10^{10}$ yr of the kicked halo neutron stars in the case of the core radius, $r_n = 6$ kpc, and the initial radius of the neutron star halo, $r_{max} = 500$ kpc. **b** The dipole moment $< \cos\theta >$, the quadrapole moment $< \sin^2 b >$, and $< V/V_{max} >$ plotted against the sampling distance for $r_n = 100$ kpc and $r_{max} = 500$ kpc. **c** The coordinate independent dipole moment, Rayleigh-Watson statistic $W$ (Briggs 1993) for randomly selected 260 orbits in the same case with **b**. **d** The coordinate independent quadrapole moment, Brighton statistic $B$. Fishman et al. (1994) found that $W = 1.1 \pm 0.3$ and $B = 6.6 \pm 1.2$ for 260 BATSE 1B burst samples.

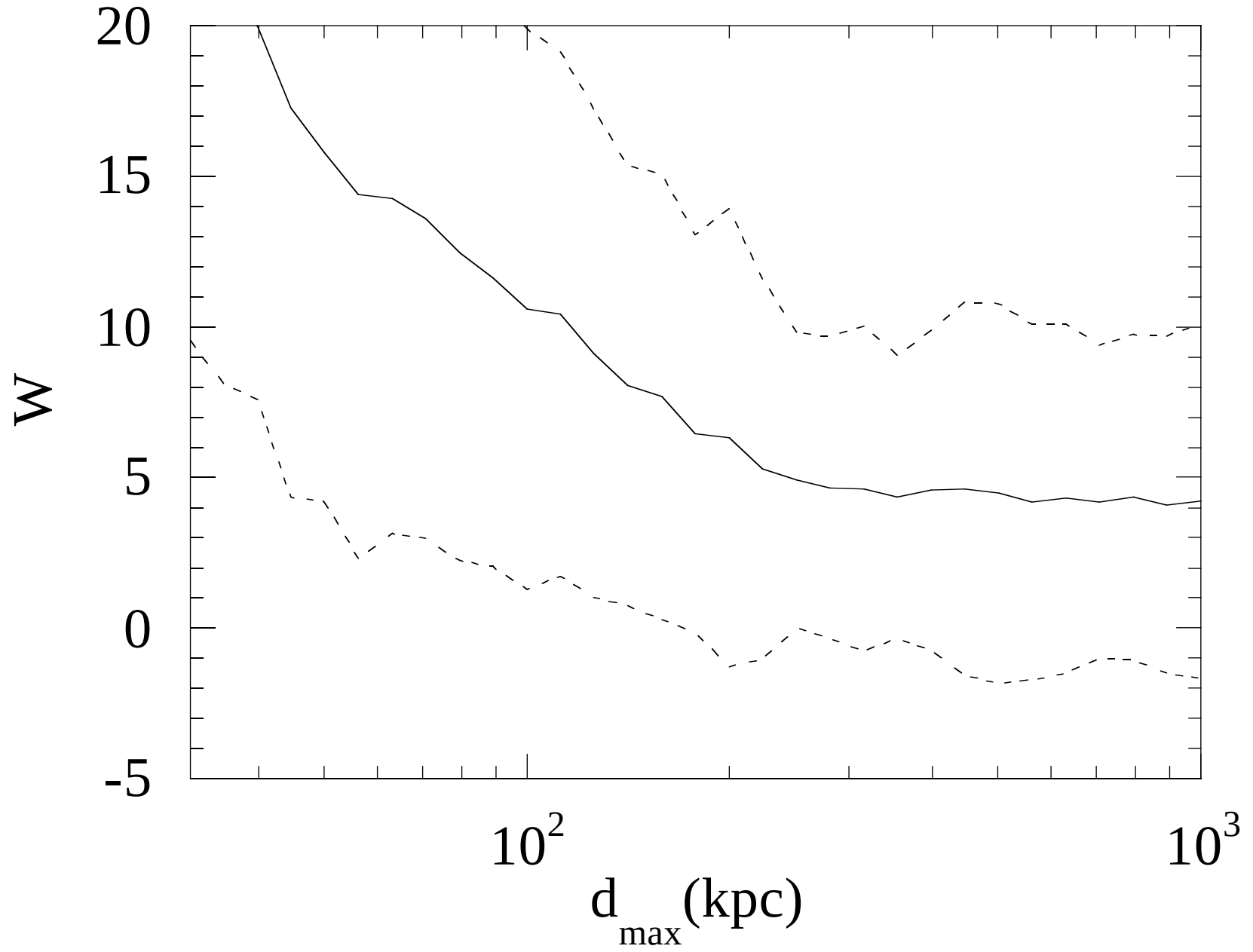

Fig. 4c

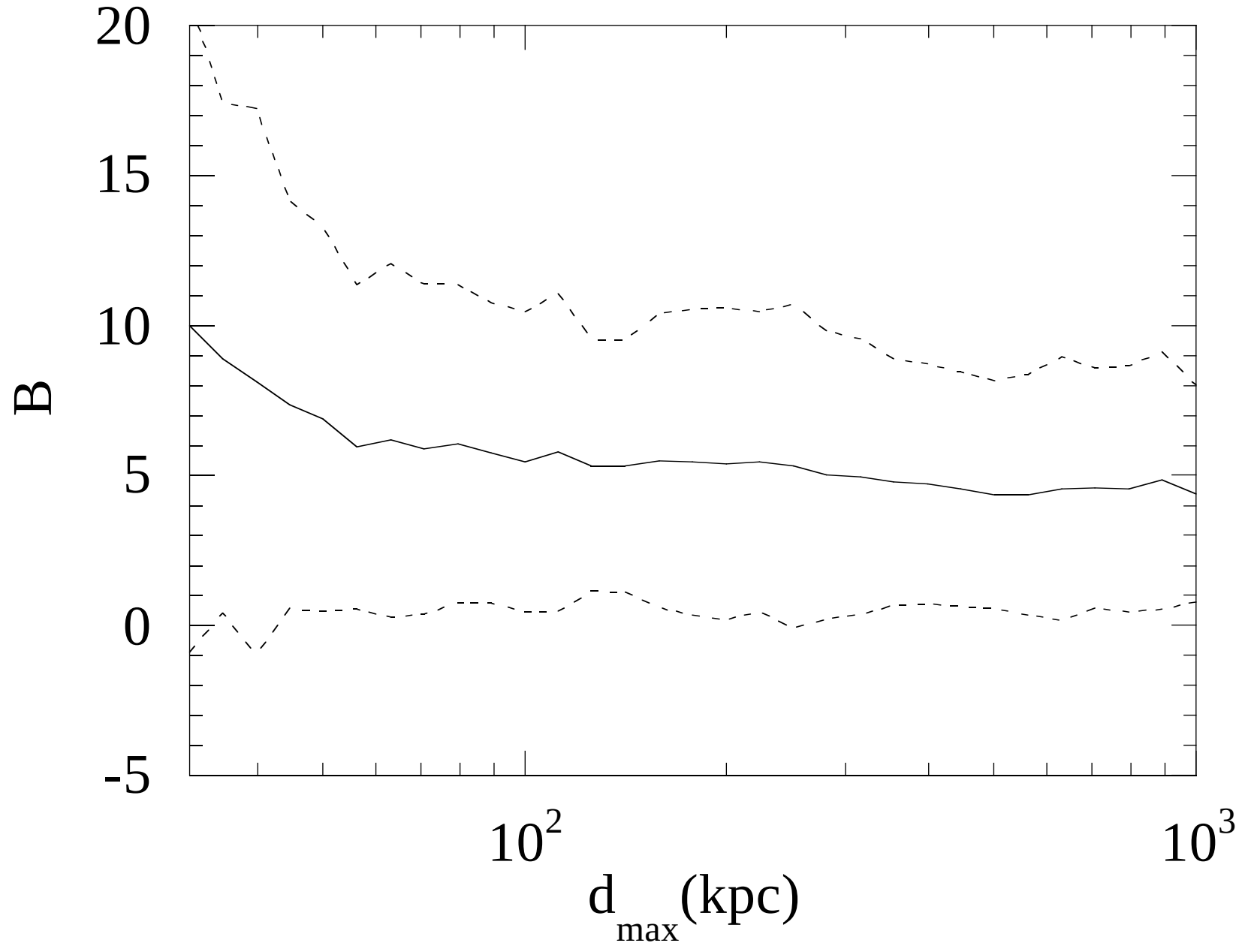

Fig. 4d

the BATSE observations. However, the fact that, on average, older pulsars have lower velocities than younger ones seems to contradict Markwardt and Ögelman's suggestion though this tendency may be due primarily to a selection effect (Lyne & Lorimer 1994).

The halo neutron star model of $\gamma$-ray bursters is attractive since it is consistent with the observation of cyclotron absorption lines and it has various theoretical and observational support from the view points of the formation and chemical evolution of galaxies (e.g. Hattori & Terasawa 1993). However the distribution of neutron stars should be fairly uniform in order to be consistent with the observed isotropy of the bursts. If we assume a core-halo structure for the burster distribution, the core radius($r_c$) should be much larger than the distance of the sun from the Galactic center ($r_c \gg$ 8.5kpc) and the sampling distance of the bursts must be even larger to be consistent with the non-uniformity of the bursters implied by the value of $< V/V_{max} >$. We may be able to expect that the kick to neutron stars can extend the core radius and the distribution of the bursters to be consistent with the observations even if we assume the initial distribution of the neutron stars has the same structure as the standard dark matter halo of the Galaxy and similar core radius. In order to examine the effect of the kick velocity to the halo neutron stars on their distribution, we traced the orbits of the halo neutron stars kicked according to the kick velocity distribution of radio pulsars. Contrary to our expectation, the core radius of the neutron star distribution is not extended by the kick velocity (Fig. 4a) when we assume a core-halo structure similar to the Galactic dark matter for the initial neutron star distribution. This result indicates that the initial core radius of the neutron star halo must be large enough compared with the distance between the Sun and the Galactic center and the sampling distance must be still larger than the core radius if we assume a core-halo structure for the initial neutron star distribution. The halo neutron star model of the $\gamma$-ray bursters is, however, still consistent with observations when we take the initial distribution of halo neutron stars appropriately, and we need several orders of magnitude more samples to exclude the extended halo model (Fig. 5d, e).

In the halo model, disk origin neutron stars must be contained in the observed burst samples if the nature of the halo origin neutron stars and the disk origin neutron stars are not different from the burster sources. The disk origin bursters make the statistics worse as shown in the above and the halo origin bursts must dominate the number of bursts. If the sampling distances are same for the disk origin bursters and the halo origin neutron stars, the ratio of the number of halo bursters to that of disk bursters should be larger by at least 2 orders of magnitude. If the type-II supernova rate is 0.01yr$^{-1}$, this means the total number of the halo origin neutron stars is at least around $10^{10}$ and their total mass exceeds 10% of the luminous mass of the Galaxy.

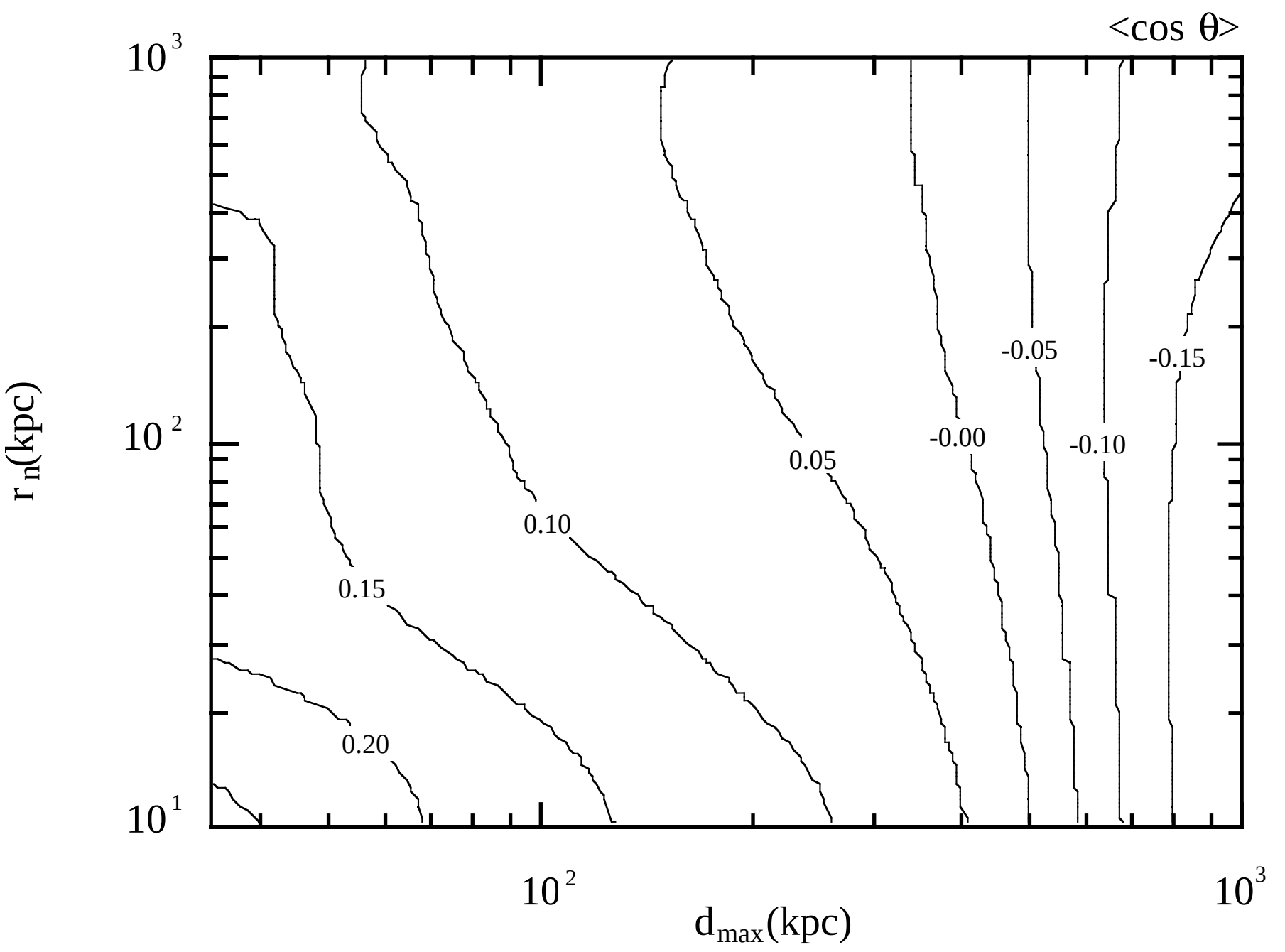

Fig. 5a

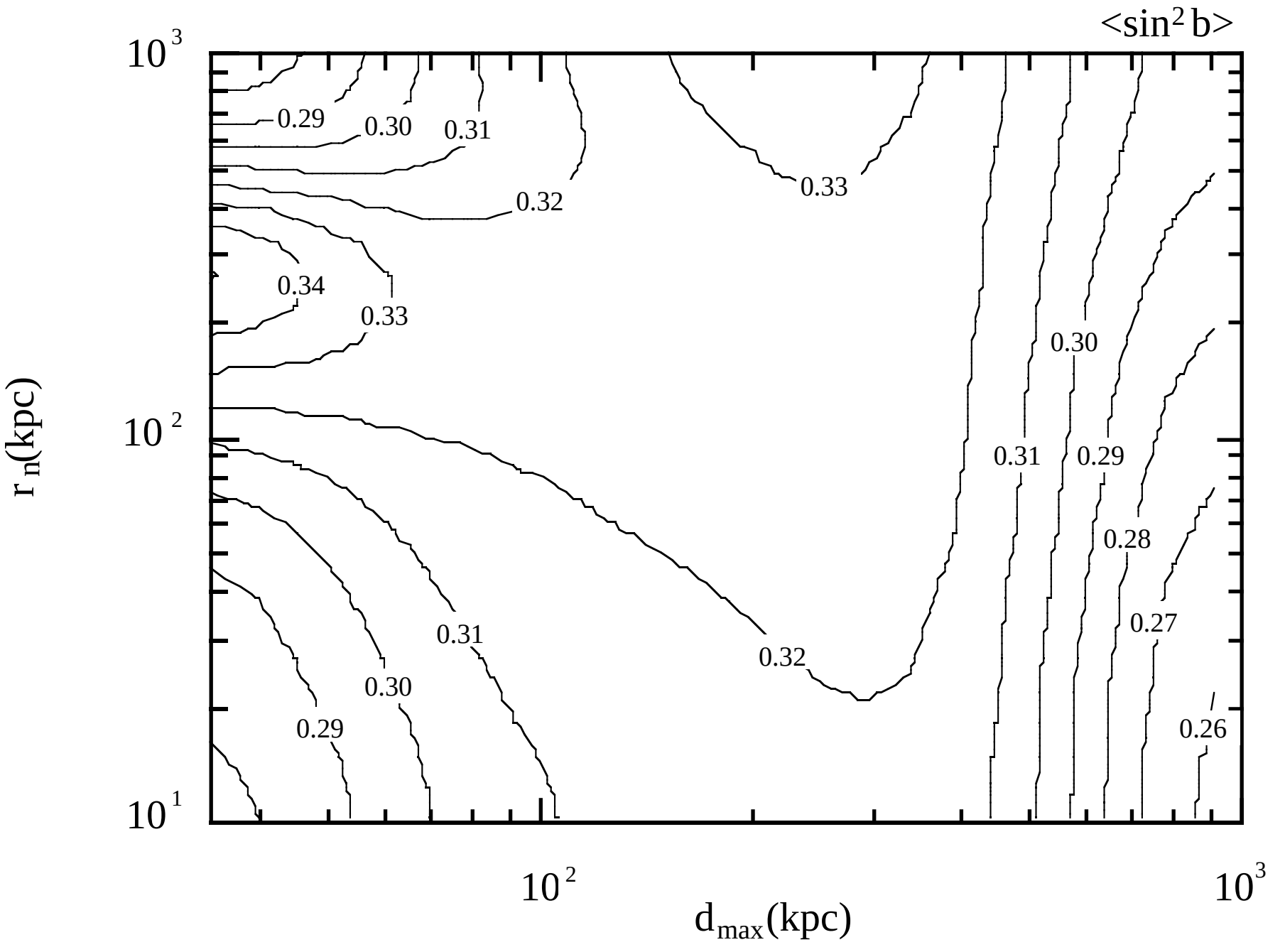

Fig. 5b

**Fig. 5a–e** Constraints on the kicked halo neutron star model defined by core radius ($r_n$) and sampling distance ($d_{max}$). **a** $< \cos\theta >$. **b** $< \sin^2 b >$. **c** $< V/V_{max} >$. **d** 90% confidence levels of the model, the solid line shows the 90% confidence level from $< \cos\theta >$ and $< \sin^2 b >$, and the broken line shows the 90% confidence level from $< \cos\theta >$, $< \sin^2 b >$, and $< V/V_{mx} >$. **e** The number of bursts which will be need to shrink the parameter space with 90% confidence.

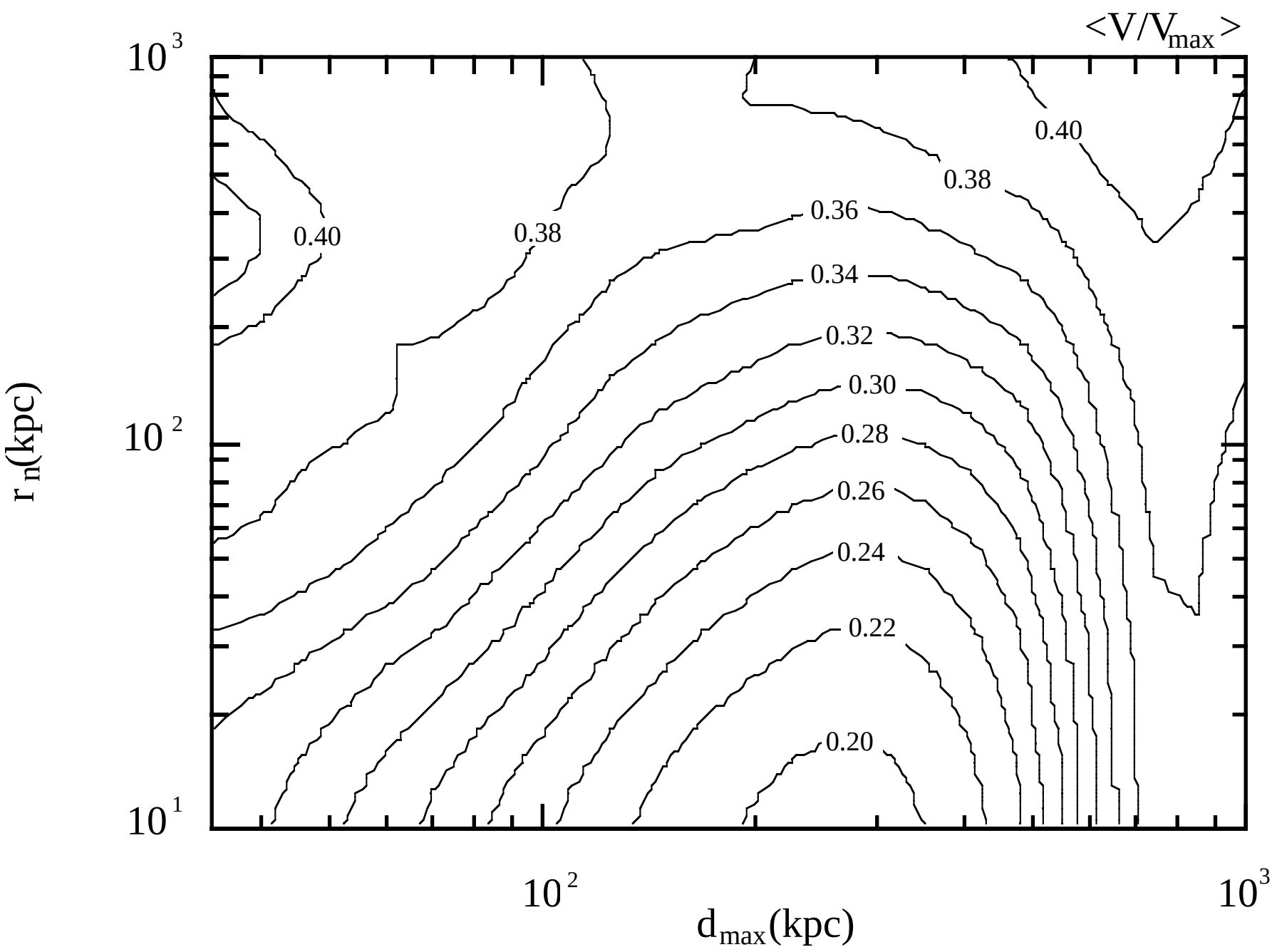

Fig. 5c

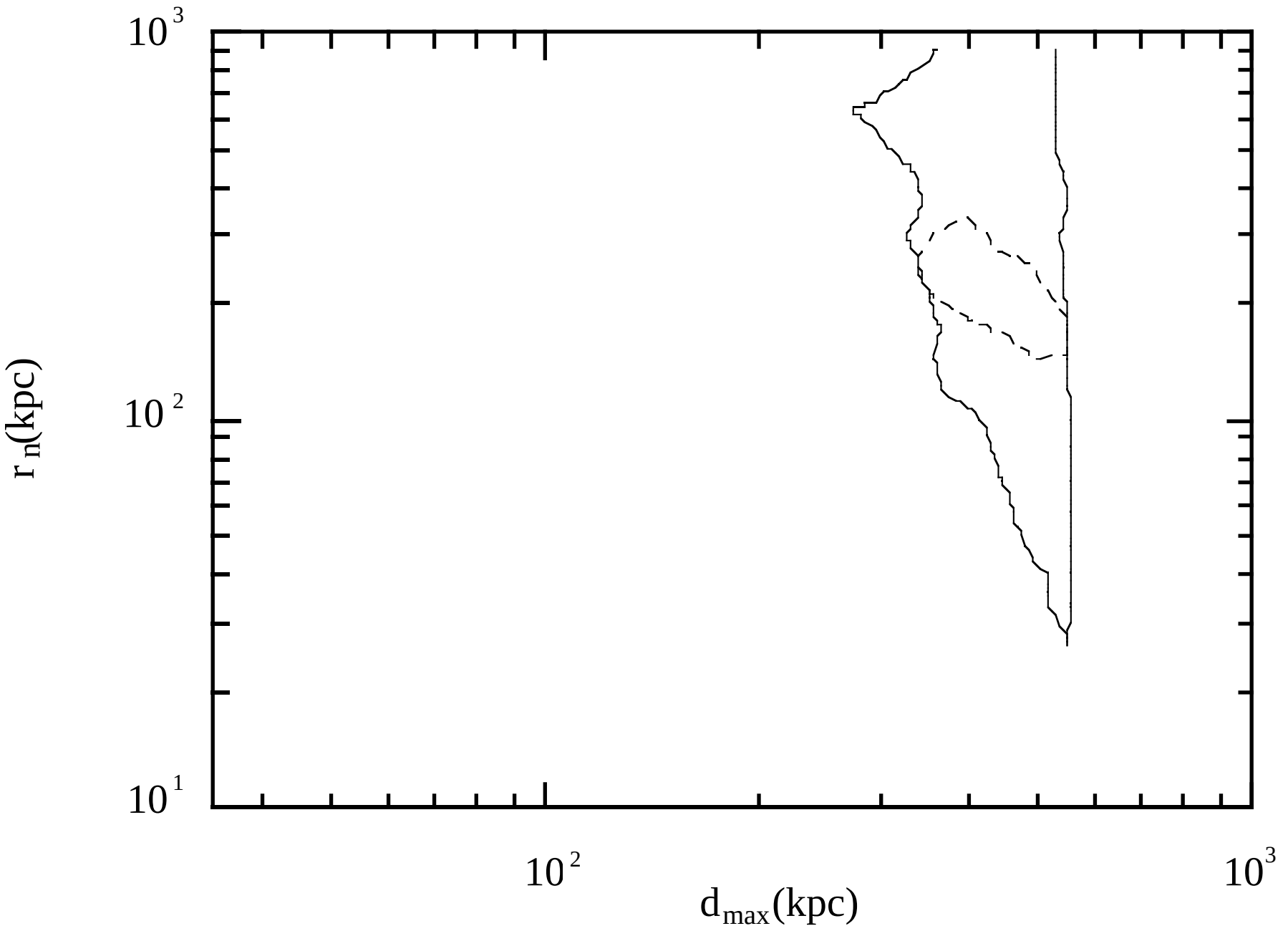

$10^3$
$10^2$
$10^1$
$r_n$(kpc)
$d_{max}$(kpc)

Fig. 5d

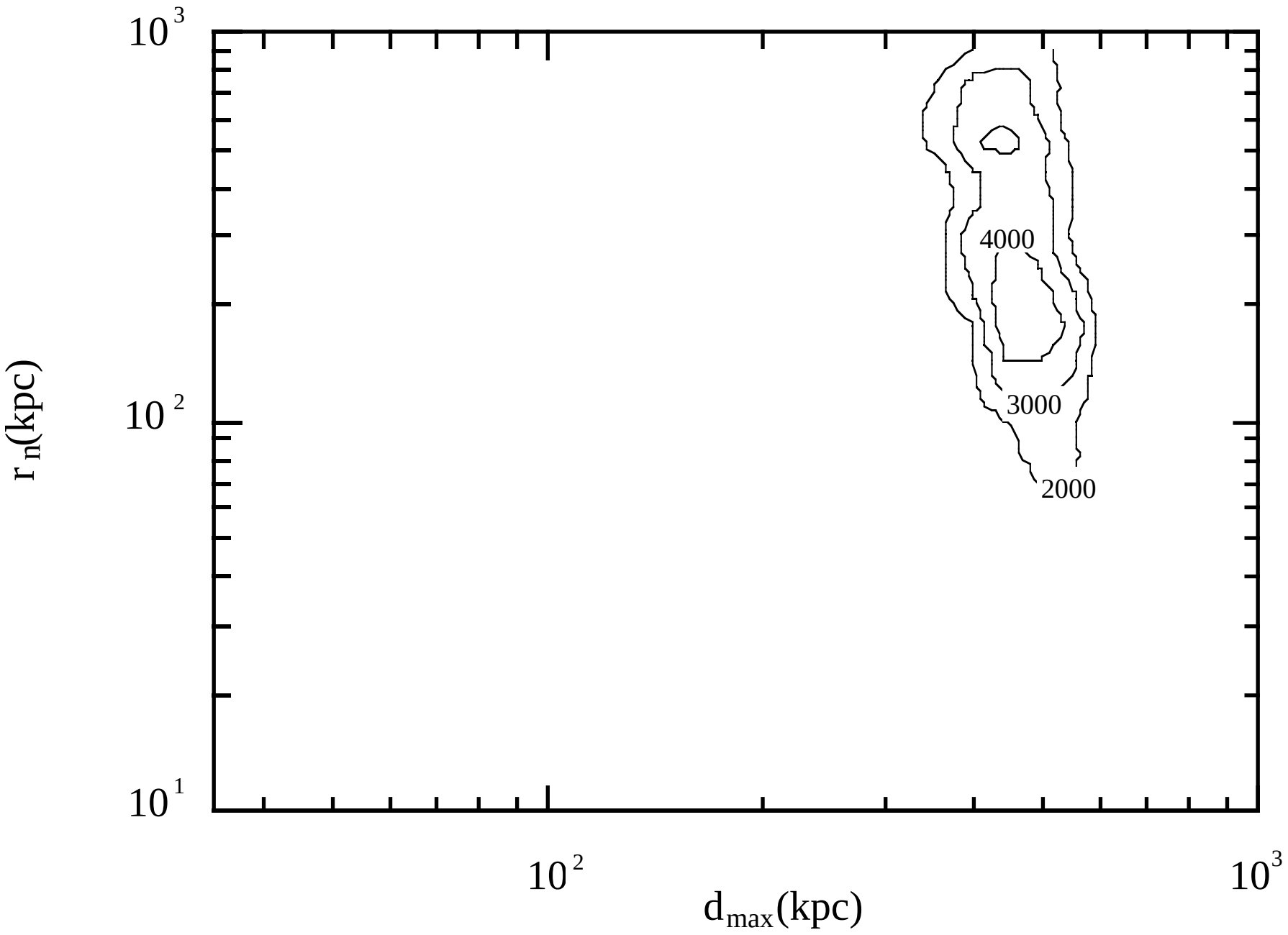

Fig. 5e

## 4 Conclusions and discussion

In conclusion, the kick velocity to neutron stars do not improve the statistics both of the disk origin model and the halo origin model of the $\gamma$-ray bursters.

1. The disk neutron star model can not be consistent with all the statistics for the spatial distribution of the $\gamma$-ray bursters simultaneously if all the neutron stars are potential *gamma*-ray bursters. The dipole moment and the quadrapole moment of the disk neutron stars are consistent with those of $\gamma$-ray bursters when the sampling distance is short enough, ie $d_{max}\lesssim 1$ kpc. However the value of $< V/V_{max} >$ for disk neutron stars is $\sim 0.5$ in this case and is clearly inconsistent with observations of $\gamma$-ray bursters.

2. If only the high velocity neutron stars with the spatial velocity greater than 750 km s$^{-1}$ can become $\gamma$-ray bursters, the model is consistent with observations in the range of the sampling distance 100 kpc $\lesssim d_{max} \lesssim$ 400 kpc with a 90% confidence level.

3. The halo neutron star model is consistent with the spatial distribution of $\gamma$-ray bursters. The core radius must be extraordinarily large ($r_n \gtrsim$ 100 kpc) and the sampling distance must be even larger if a core-halo structure is assumed for the initial distribution of halo neutron stars.

Each Galactic neutron star model thus suffers from some difficulties or unnaturalness. The simplest disk neutron star model cannot explain all the statistics at the same time as described above. If the sampling distance is, however, small enough as $d_{max} \lesssim 1$ kpc, the dipole moment and the quadrapole moment of the local disk neutron stars can be consistent with observations. This model benefits from the inclusion of the kick velocity since the local structure in the stellar distribution is erased and uniform distribution of the bursters is realized. However the model needs an additional component of the bursters in order to be consistent with the observed deviation from uniformity of the burster distribution (e.g. Smith & Lamb 1993) unless the slight deviation from uniformity results in the observed small value of $< V/V_{max} >$ or the result of $< V/V_{max} >$ is an experimental artifact. The luminosity of a typical $\gamma$-ray burster at distance $d$ is estimated to $\sim 10^{35-36}(d/100\mathrm{pc})^2$ erg s$^{-1}$. This implies the released energy by a typical burster is $\sim 10^{36-37}(d/100\mathrm{pc})^2$ erg, which can be easily obtained, for instance, from thermonuclear burning of accreted matter onto the neutron star surface. The recurrence rate of the bursts in the model is short enough to observe $\sim 0.1 - 10\mathrm{yr}^{-1}$.

The difficulty in the fast moving neutron star hypothesis is the lack of a natural physical mechanism to suppress slowly moving neutron stars to become $\gamma$-ray bursters. The implied burst luminosity in this model is of the order $10^{41-42}$ erg s$^{-1}$. This corresponds to a burst energy of order $\sim 10^{42-43}$ erg, which is though to be available, for instance, from starquakes such as crustquakes, glitches and so on. Li and Dermer (1992) have estimated the burst recurrence rate to be $10^{-3}$ yr$^{-1}$.

The problem in the halo neutron star model is how to realize an extended neutron star halo in a Galactic formation scenario. The implied extraordinarily large core radius is quite different from that of the dark matter halo of the Galaxy and we do not know so far any objects with such a distribution. The released energy by bursters in this model is of the same order as that of the fast neutron star model $\sim 10^{42-43}$ erg, and the recurrence rate is estimated to $10^{-7}$ yr$^{-1}$ from, for instance, metal enrichment of the intracluster medium ( Hattori & Terasawa 1993).

One way to save the Galactic neutron star model from its associated difficulties would be to suppose that $\gamma-$ray bursters are old neutron stars which were accelerated to velocities faster than 750 km s$^{-1}$ by jet propulsion and passed the death line for pulsars due to the spin down caused by the rotational energy loss by jet ejection (Markwardt & Ögelman 1995).

Another possibility is the bimodal star formation model for the Galaxy in which the initial star formation burst of the Galaxy is assumed to have occurred fairly uniformly in the extended halo region. An extended and uniform halo of neutron stars as the remnants of the initial star formation burst may be formed in this manner. Such a model is supported by the chemical evolution of clusters of galaxies (Hattori and Terasawa 1993).

The $< V/V_{max} >$ test applied to the BATSE observation is usually interpreted that it has confirmed inhomogeneity of the $\gamma$-ray burster distribution. A $< V/V_{max} >$ test is, however, an indirect method to examine homogeneity of sources, and we can not necessarily rule out the possibility that $< V/V_{max} >$ is biased by detector thresholds.

If the restriction of the source inhomogeneity is nothing but an experimental ghost, the classical local disk model can not be excluded. We may need to wait for next generation satellite experiments to fix this problem.

A test for the Galactic neutron models would be the observations of recurrences of bursts which is expected in the local disk neutron star model. From BATSE observations it is still unclear whether or not $\gamma$-ray bursts repeat. However, it would lend strong support for Galactic models if cyclotron absorption lines are observed fairy universally in the $\gamma$-ray burster spectra. The HETE satellite is expected to be able to settle both problems.

## 5 Acknowledgement

The authors are grateful to Ian Bond for valuable comments.